\documentclass[12pt,epsfig]{article}
\usepackage{color}
\usepackage{amssymb}
\usepackage{amsmath}
\usepackage[dvips]{graphicx}
\usepackage{bm}
\usepackage{multirow}
\usepackage{rotating}
\usepackage{amsmath}
\begin{document}
\begin{center}
\textbf {Use of spatial cross correlation function to study formation mechanism of massive elliptical galaxies}\\
\vskip.2in Tuli De$^{1}$, Tanuka Chattopadhyay$^{2}$\\
{\small and}\\
Asis Kumar Chattopadhyay$^{3}$\\
\vskip.1in
$^{1}$Department of Mathematics, Heritage Institute of Technology, Kolkata\\Choubaga Road, Anandapur, Kolkata -700107, India\\
email: tuli.de@heritageit.edu\\
 \vskip.1in
$^{2}$Department of Applied Mathematics, Calcutta University, Kolkata, India\\
92 A.P.C. Road, Kolkata -700009\\
email: tanuka@iucaa.ernet.in\\
\vskip.1in
$^{3}$Department of Statistics, Calcutta University, Kolkata\\35 Ballygunge Circular
 Road,Kolkata-700019, India\\
email: akcstat@caluniv.ac.in
\end{center}

\begin{abstract}
Spatial clustering nature of galaxies have been studied previously through auto correlation function.
The same type of cross correlation function has been used to investigate parametric clustering
nature of galaxies e.g. with respect to masses and sizes of galaxies.\\
Here formation and evolution of several components of nearby
massive early type galaxies $( M_{*} \geq 1.3\times 10^{11}
M_{\bigodot})$ have been envisaged through cross correlations, in
the mass-size parametric plane, with high redshift $( 0.2 \leq z
\leq 7)$ early type galaxies (hereafter ETG).It is found that the
inner most components of nearby ETGs have significant correlation
$(\sim 0.5\pm (0.02-0.07))$ with ETGs in the highest redshift range
$(2\leq z\leq7)$ called $`$red nuggets' whereas intermediate
components are highly correlated $(\sim 0.65\pm (0.03-0.07))$ with
ETGs in the redshift range $0.5 \leq z \leq 0.75$. The outer most
part has no correlation in any range, suggesting
a scenario through in situ accretion.\\
The above formation scenario is consistent with the previous results
obtained for NGC5128 (Chattopadhyay et al. (2009); Chattopadhyay et
al. (2013)) and to some extent for nearby elliptical galaxies (Huang
et al. (2013)) after considering a sample of ETGs at high redshift
with stellar masses greater than or equal to
$10^{8.73}M_{\bigodot}$. So the present work indicates a three phase
formation of massive nearby elliptical galaxies instead of two as discussed in previous works.
\end{abstract}

\vspace{0.1in}\textbf{Keywords:}: Cross correlation, Elliptical galaxies, Clustering.\\

 \section{Introduction}In 1934 Hubble observed that the frequency distribution of the count of galaxies over
 the space is strongly skew but the distribution of its logarithm is close to symmetric. Bok (1934) and Mowbray (1938)
 found that variance of the count is considerably larger than expected for a random galaxy distribution.
 Such studies indicates that locally galaxies are clustered over space. Several attempts have been made to study
 this clustering nature on the basis of angular positions of the galaxies. Most of them (Zwicky(1953),
 Limber (1953,1954), Chandrasekhar and Munch (1952)) have used spatial and angular correlation functions
 to study this phenomenon. In this area the contributions of Neyman and Scott(1954) is very significant.
 This spatial clustering nature motivated us to investigate the clustering nature with respect to the other
 parameters also by using the same approach.\\
Classical formation of elliptical galaxies can be divided into
five major categories
 e.g. (i) the monolithic collapse model, (Larson (1975); Carberg (1984); Arimoto and Yoshii (1987)) (ii) the major merger
 model (Toomre (1977); Ashman and Zepf (1992); Zepf et al. (2000); Bernardi et al. (2011); Prieto et al. (2013)),
 (iii) the multiphase dissipational collapse model (Forbes (1997)), (iv) the dissipationless merger model
 (Bluck et al. (2012); Newman et al. (2012)) and (v) the accretion and in situ hirarchical merging (Mondal et al. (2008)).
  Recent observations in the deep field have explored that high redshift galaxies have the size of the order of
 1 kpc (Daddi et al. (2005); Trujilo et al. (2006); Damjanov et al. (2011)) and have higher velocity dispersion
 (Cappellari et al. (2009); Onodera et al. (2009)) than nearby ETGs of the same stellar mass. Galaxies at intermediate
 redshifts (since $z\approx 2.5$) have stellar masses and sizes increased by a factor almost 3-4
 (Van Dokkum et al. (2010); Papovich et al. (2012); Szomoru et al. (2012)). All these evidences suggest that massive ETGs
 form in two phases viz. inside -out i.e. intense dissipational process like accretion (Dekel et al. (2009)) or major
 merger form an initially compact inner part. After this a second slower phase starts when the outer most part is developed
 through non-dissipational process e.g. dry, minor merger. The above development arising both in the field of observations
 as well as theory, severely challenge classical models like monolithic collapse or major merger and favors instead a
 $``$two phase" scenario (Oser et al. (2010); Johanson et al. (2012)) of the formation of nearby elliptical galaxies. The task
 remains, is to check whether the compact inner parts of the nearby ETGs have any kind of similarity with the fossil bodies
 (viz. $`$red nuggets') at high redshift.\\
 In a previous work (Huang el al. (2013)), the authors have pursued the above task through matching $`$median' values of the two
  systems. They used this measure with respect to univariate data and the univariate data they considered, are either stellar $`$mass'
   or $`$size'. For ETGs in the redshift range, $0.5\leq z\leq7$, considered, in the present data set, the stellar mass-size
   correlation is $r(M_{*},R_{e}) = 0.391$, p-value=0.00. For nearby ETGs for inner, intermediate and outer components the stellar
   mass-size correlations with p-values are $r(M_{*},R_{e}) = 0.720$, p-value=0.00, $r(M_{*},R_{e}) = 0.636$, p-value=0.00 ,$r(M_{*},R_{e}) = 0.573$,
   p-value=0.00 respectively and all these values are highly significant.
 Hence use of univariate median matching is not sufficient in the present context for highly correlated bivariate data. Also, median does not
 include all objects in a particular data set. For this a more sophisticated technique is in demand for such kind of investigation.\\
 In the present work we have used the mass-size data of high red shift galaxies and nearby ETGs and used a cross-correlation, especially
 designed to study bivariate data. This is more trustworthy and meaningful in the present situation. In section 2 we have discussed the
 data set and in section 3 we have described the method. The results and interpretations are given under section 4.

\section{Data sets:}
we have considered eight data sets. Data sets 1-3 consist of stellar
masses and sizes of 70 nearby ETGs taken from Ho et al. (2011).
There are three components corresponding to each massive ETG,
described by a single S$\acute{e}$rsic (1968) index, as considered
by Huang et al.(2013). They are, (i)an inner component with
effective radii $R_{e} \leq 1$ kpc, (ii)an intermediate component
with effective radii $R_{e} \sim 2.5 $ kpc and (iii) an outer
envelope with $R_{e}\sim 10$ kpc. Data sets 4 -8 consist of stellar
masses and effective radii of high redshift ETGs with stellar masses
$M_{*}\geq 10^{8.73}M_{\bigodot}$ in the redshift bins $0.5 < z \leq
0.75, 0.75 < z\leq 1, 1 < z \leq 1.4, 1.4 < z \leq 2.0, 2.0 < z \leq
2.7$. Unlike Huang et al. (2013) we also included
intermediate mass high redshift galaxies. Data sets 4-8 contains 786 high redshift ETGs from the following works.\\
392 galaxies $(0.2 \leq z \leq 2.7)$ from Damjanov et al. (2011), 32
$(1.5 < z < 3)$ galaxies from GOODS-NICMOS survey (Conselice et al.
(2011)) for S$\acute{e}$rsic (1968)index n $>$ 2, 21 galaxies from
CANDELS (Grogin et al. 2011) $(1.5 < z < 2.5)$, 107 from Papovich et
al. (2012)$(1.5 \leq z \leq 2.5)$, 48 from Mclure et al. (2012)($1.3
< z < 1.5$), 62 from Saracco et al. (2011) $(1 < z < 2)$, 124 galaxies from Nilsson et al. (2013).\\
Since the data sets are chosen from different sources, they have
various selection biases and errors etc. Hence, to judge their
compatibility we have performed a  multivariate multi sample
matching test (Puri \& Sen (1966): Appendix, Mckean (1974)) to see
whether they have the same parent distribution or not. From
previous works it is evident that galaxies have undergone
cosmological evolution via merger or accretion (Naab (2013);
Khochfar $\&$ Silk (2006); De Lucia $\&$ Blaziot (2007); Guo $\&$
White (2008); Kormendy et al. (2009); Hopkins et al. (2010)) and
we have performed the matching test for galaxies within the same
redshift zone. The data set taken from Damjanov et al. (2011)
contains maximum number of galaxies within the entire redshift
zone ($0.2 \leq z \leq 2.7$) used in the present analysis. For
this we have compared it with the other sets. The results are
given in Table 1. It is clear from Table 1 that all the tests are
accepted except one (Papovich et al. (2012)) where the matching
redshift zone is very narrow. Since almost in 99$\%$ cases the
test is accepted we assume that the dataset consisting of samples
from different sources is more or less
homogeneous with respect to mass-size plane.\\
 It is to be noted that in Ho et
al. (2011) paper, the magnitude values of the three components
corresponding to each ETG are given from which, luminosities are
computed. Then these luminosities are multiplied by $(M/L)$ ratios
for obtaining stellar masses. The $(M/L)$ ratios are computed
following Bell et al. (2001). we have not become able to retrieve
data for some high redshift galaxies and instead included some new
data from other recent references so that sample size of high
redshift galaxies are some what reduced in our case, but the overall
distribution of these galaxies are similar in the size-redshift
plane with those considered by Huang et al. (2013a) (viz. Fig.1 of
Huang et al. (2013a) and Fig.1 in the present work) except the region
$1 \leq z \leq 2 $ which is more populated than Huang et al. (2013a)
sample as we have included new galaxies in data sets 4-8.

\section{Method:} The theory of the spatial distribution of galaxies has been discussed by several authors
like Peebles (1980), Blake et al. (2006), Martinez and Saar (2012)
etc. During 1950s, the most extensive statistical study was
performed by Neyman and Scott. Their work was based on the large
amount of data obtained from the LICK survey. The main empirical
statistics they used were the angular auto correlation function of
the galaxy counts(Neyman et al. (1956)) and Zwicky's index of
clumpiness
(Neyman et al. (1954)).\\
Neyman and Scott (1952) introduced this theory on the basis of four
assumptions viz. (i) galaxies occur only in clusters, (ii) The
number of galaxies varies from cluster to cluster subject to a
probabilistic law, (iii) the distribution of galaxies within a
cluster is also subject to a probabilistic law and iv) the
distribution of cluster centres in space is subject to a
probabilistic law described as quasi-uniform. As the observed
distribution of number of galaxies does not follow Poisson law, it
is suspected that not only the apparent but also the actual
spatial distribution of galaxies is clustered.\\
In the present work attempts have been made to establish the same
postulates with respect to mass-size distribution of galaxies. Here
the hypothesis is $``$there is clustering nature also in the galaxy
distribution with respect to the parameters mass and size of the
galaxies". This particular hypothesis also has been studied by
several authors. But we have followed the same approach
as that used to establish spatial clustering as discussed above.\\
In cosmology the cross correlation function $\xi(r)$ of a
homogeneous point process is defined by (Peebles (1980))
\begin{equation}
dP_{12}=\pi^{2}[1+\xi(r)]dV_{1}dV_{2}
\end{equation}
where r is the separation vector between the points $x_{1}$ and
$x_{2}$ and $\pi$ is mean number density.Considering two
infinitesimally small spheres centered in $x_{1}$ and $x_{2}$ with
volumes $dV_{1}$ and $dV_{2}$,the joint probability that in each of
the spheres lies a point of the point process is
\begin{equation}
dP_{12}=\lambda_{2}(x_{1},x_{2})dV_{1}dV_{2}
\end{equation}
In (2),$\lambda_{2}(x_{1},x_{2})$ is defined as the second order
intensity function of a Point process.\\
If the point field is homogeneous, the second-order intensity
function $\lambda_{2}(x_{1},x_{2})$ depends only on the distance
$r=|x_{1}-x_{2}|$ and direction of the line passing through $x_{1}$
and $x_{2}$. If, in addition, the process is isotropic,the direction
is not relevant and the function only depends on r and may be
denoted by $\lambda_{2}(r)$. Then
\begin{equation}
\xi(r)=\frac{\lambda_{2}(r)}{\pi^{2}}-1
\end{equation}
Different authors proposed several estimators of $\xi(r)$. Natural
estimators have been proposed by Peebles and Hauser (1974). The
cross correlation function $\xi(r)$ can be estimated from the galaxy
distribution by constructing pair counts from the data sets. A pair
count between two galaxy populations 1 and 2, $D_{1}D_{2}(r)$, is a
frequency corresponding to separation r to r+$\delta$r for a bin of
width $\delta$r in the histogram of the distribution r, $D_{i}R_{j}$
and $R_{i}R_{j}$ denote the same pair count corresponding to one
galaxy sample and and one simulated sample and two simulated samples respectively, i, j=1, 2. Two
natural estimators are given by
\begin{equation}
\hat{\xi_{1}}=\frac{D_{1}D_{2}(r)}{D_{2}R_{1}(r)}-1
\end{equation}

\begin{equation}
\hat{\xi_{2}}=\frac{D_{1}D_{2}(r)}{D_{1}R_{2}(r)}-1
\end{equation}
Another two improved estimators are(Blake et al(2006))
\begin{equation}
\hat{\xi_{3}}=\frac{D_{1}D_{2}(r)R_{1}R_{2}(r)}{D_{1}R_{2}(r)
D_{2}R_{1}(r)}-1
\end{equation}
and
\begin{equation}
\hat{\xi_{4}}=\frac{D_{1}D_{2}(r)-D_{1}R_{2}(r)-D_{2}R_{1}(r)+R_{1}R_{2}(r)}{R_{1}R_{2}(r)}
\end{equation}
The first two estimates are potentially biased. As in the present
situation we are considering mass-size parametric space, we have
taken r as the Euclidean distance between two (mass,size) points of
two galaxies either original or simulated. In order to generate
simulated samples of mass and size, we have used uniform
distribution of mass and size with ranges selected from original
samples. Here r is normalized by dividing the original separation by
the maximum separation.The variances of the estimators are measured by
bootstrap method.

\section{Results and discussion:} We have computed the cross-correlation functions of each of data
sets 1-3 with data sets 4-8 i.e. we have tried to find any kind of
correlation between three components of nearby ETGs with high
redshift ETGs in five redshift bins as mentioned above. We have
found significant
 correlation between data set 1 and data set 8 and between data set 2 and data set 4. This is clear from
 Figs. 2 and 3 respectively where the correlations are as high as $0.5\pm(0.02-0.07)$ and $0.65\pm(0.03-0.07)$
 for both the estimates at minimum separation $(viz. r \sim 0.1)$. These show that the innermost components
 of nearby elliptical galaxies are well in accordance with highest redshift massive ETGs (viz.
 mass $\sim $ $10^{11.14}M_{\bigodot}$ and $R_{e}$$\sim$ 0.92 kpc), known as $`$red nuggets' but the
 intermediate components are highly correlated with galaxies in the redshift bin $0.5 \leq z \leq 0.75$ having
 median mass and size, $10^{10.87}M_{\bigodot}$ and 2.34 kpc respectively. If we merge data sets 1 and 2 and
 compare with high z galaxies in five redshift bins, the cross-correlation functions are all close to zero at
 all separations unlike Huang et al (2013a).\\
 The above result is somewhat consistent with the work of Huang et al. (2013a) in a sense that the inner and intermediate parts
are the fossil evidences of high red shift galaxies but unlike Huang
et al. (2013a), components 1 and 2 together show no correlation with
all high redshift ETGs together they are highly correlated with high
redshift ETGs in two red shift bins and this indicates clearly two
different epochs of structure formation as shown by their z values.\\
After finding the cross correlation functions between data sets 1 and 8, we have fitted a power law assuming the relation \\
\begin{equation}
\xi(r) \propto \frac{1}{r}
\end{equation}
i.e.,
\begin{equation}
\xi(r) = A r^{-1}
\end{equation}
Where for estimator 1, A =0.02672 and for estimator 2, A=0.031395.\\
We have also performed Kolmogorov Smirnov test for justifying the
goodness of fit of the power law. Here we have assumed that the
cross correlations and the fitted values are samples coming from
the distribution function of a Pareto distribution.The p-values
for this test for estimator 1 is 0.4175 and for estimator 2 is
0.7869, signifying that the tests are accepted and the fitted
power law gives well justification for the cross correlation and
distance relationship.We have fitted similar power laws for cross
correlation function and distance for data sets 2 and 4. Here the
proportionality constants are 0.0247 for estimator 1 and 0.0369
for estimator 2. The Kolmogorov Smirnov tests give p-values =
0.7869 for both the estimators, signifying that in this case also
the relationship is well justified.\\
On the other hand, cross-correlation function, computed for data set 3 with galaxies in the above five bins are all
insignificant which is clear from Fig. 4.\\
During the formation of massive ellipticals,major and minor merger
play a significant role for the morphological and structural
evolution (Naab (2013); Khochfar $\&$ Silk (2006); De Lucia $\&$
Blaizot (2007); Guo $\&$ White (2008); Kormendy et al. (2009);
Hopkins et al. (2010)). The most massive elliptical galaxies (or
their progenitors) are considered to start their evolution at
z$\sim$ 6 or higher in a dissipative environment and rapidly become
massive($\sim 10^{11}M_{\bigodot}$) and compact at z$\sim$2 (Dekel
et al. (2009); Oser et al. (2010); Feldmann et al. (2011); Oser et
al. (2012)). Also a significant fraction is observed to be quiescent
at z$\sim$2, 4-5 times more compact and a factor of two less massive
than their low redshift descendants( van Dokkum et al. (2008); Cimatti et al.
(2008); Bezanson et al. (2009); van Dokkum et al. (2010); Whitaker
et al. (2012)). Now,for the massive ellipticals in the present
sample, the innermost cores (data set 1) are well in accordance with
highest redshift ($2.0< z\leq 2.7$) galaxies and their core masses
(viz. median value $\sim 10^{10.203}M_{\bigodot}$ and
$10^{10.6839}M_{\bigodot}$ respectively). Hence it is reasonable to
separate that these high redshift population forms the cores of at
least some, if not all, present day massive ellipticals. Thus
formation of massive ellipticals only by monolithic collapse model
is challenged because they will be too small and too red(van Dokkum
et al. (2008); Ferr$\acute{e}$-Mateu et al. (2012)), the subsequent
evolution forming the intermediate (data set 2) and outer part (data
set 3)might be as follows. On the aspect of major or minor
major,following Naab et al. (2009) it is seen that if $M_{i}$ and
$r_{i}$ be the mass and radius of a compact initial stellar system
with a total energy $E_{i}$ and mean square speed $<v^{2}_{i}>$ and
$M_{a},r_{a},E_{a}$ and $<v^{2}_{a}>$ be the corresponding values
after merger with other systems then,
\begin{equation}
\frac{<v^{2}_{f}>}{<v^{2}_{i}>}=\frac{(1+\eta_{\varepsilon})}{1+\eta}
\end{equation}
\begin{equation}
\frac{r_{g,f}}{r_{g,i}}=\frac{(1+\eta)^{2}}{(1+\eta_{\varepsilon})}
\end{equation}
\begin{equation}
\frac{\rho_{f}}{{\rho}_{i}}=\frac{(1+\eta_{\varepsilon})^{3}}{(1+\eta)^{5}}
\end{equation}
where the quantities with suffix $`$f' are the final values,
$\eta=\frac{M_{a}}{M_{i}}$, $\varepsilon=<v^{2}_{a}>/<v^{2}_{i}>$,
$\rho$ is the density. Then for $\eta$=1(major merger), the mean
square speed remains same,the size increases by a factor of 2 and
densities drop by a factor of four. Now, in the present
situation,the intermediate part (data set 2) has radii (median
value $<R_{e}>\sim 2.560$ kpc) which is almost 3 times larger than
the
radii of inner part (median value$<R_{e}>\sim$ 0.6850 kpc).\\
Also in a previous work (Chattopadhyay et al. (2009); Chattopadhyay
et al. (2013)) on the brightest elliptical galaxy NGC 5128, we have
found three groups of globular clusters. One is originated in
original cluster formation event that coincided with the formation
of elliptical galaxy and the other two, one from accreted spiral
galaxy and other from tidally stripped dwarf galaxies. Hence we may
conclude from the above discussion that the intermediate parts of
massive elliptical is formed via major merger with the high redshift
galaxies in $0.5\leq z < 0.75$, whose median mass and size are
respectively $10^{10.87}M_{\bigodot}$ $\&$ 2.34 kpc respectively.\\
In the limit when $<v^{2}_{a}> << <v^{2}_{i}>$ or $\varepsilon <<
1$, the size increases by a factor of four (minor merger). In the
present case, the outermost parts of massive ellipticals have
sizes much larger (median value$<R_{e}>\sim$ 10.54 kpc)) than
innermost part. Also, median mass of this part is of the order of
$10^{10.6839}M_{\bigodot}$ which is comparable to the combined
masses of few dwarf galaxies. So, it might be suspected that the
outermost part is primarily composed of stellar components of
tidally accreted satellite dwarf galaxies.This is also consistent
with our previous works (Chattopadhyay et al. (2009);
Chattopadhyay et al. (2013)) in case of NGC 5128. Since data set 3
has no correlations with any subset of high redshift galaxies, we
cannot specifically confirm their formation epoch but we can at
most say that their formation process is different from the
innermost and intermediate part.\\
Finally we can conclude that formation of nearby massive ellipticals
have three parts, inner, intermediate and outermost, whose formation
mechanisms are different. The innermost parts are descendants of
high ETGs called $`$red nuggets'. The intermediate parts are formed by
major mergers in the redshift zone, $0.3\leq z < 0.75$. The outer envelop
might be formed by minor mergers with tidally stripped
satellite dwarf galaxies (Mihos
et al. (2013); Mondal et al. (2008); Chattopadhyay et al. (2009);
Chattopadhyay et al. (2013)). Since, the densities and velocity
dispersion values and abundances are not available with the present
data sets, so more specific conclusions can be drawn if these data
are available for massive ellipticals and satellite dwarfs. But at
this moment we can say, that since two different formation scenario
are very unlikely for the same galaxy at a particular epoch, so the
above study is indicative of a $`$third phase' of formation of the
outermost parts of massive nearby ellipticals rather than a $`$two
phase one' as indicated by previous authors.

\begin{table}
\caption{Multivariate multisample test for the matching of parent distributions
corresponding to data sets 4-8 (at 0.5 percent level of
significance)} \label{tab:test}
\begin{tabular}{llll}
\hline\noalign{\smallskip}
Sample1 & Sample2 & p-value & Decision \\
\noalign{\smallskip}\hline\noalign{\smallskip}
$Damjanov$ et al.(2011) & $Grogin$ et al.(2011) & 0.005 & Accepted\\
$,,$ & $Consclice$ et al.(2011) & 0.007& Accepted\\
$,,$ & $Nilsson$ et al.(2011) & 0.0447 & Accepted\\
$,,$ & $Mclure$ et al.(2012) & 0.003 & Accepted\\
$,,$ & $Saccaro$ et al. (2011) & 0.096 & Accepted\\
$,,$ & $Papovich$ et al.(2012) & 0.000 & Rejected\\
 \noalign{\smallskip}\hline
\end{tabular}
\end{table}

\begin{figure}
  \centering
  \includegraphics[width=0.7\textwidth]{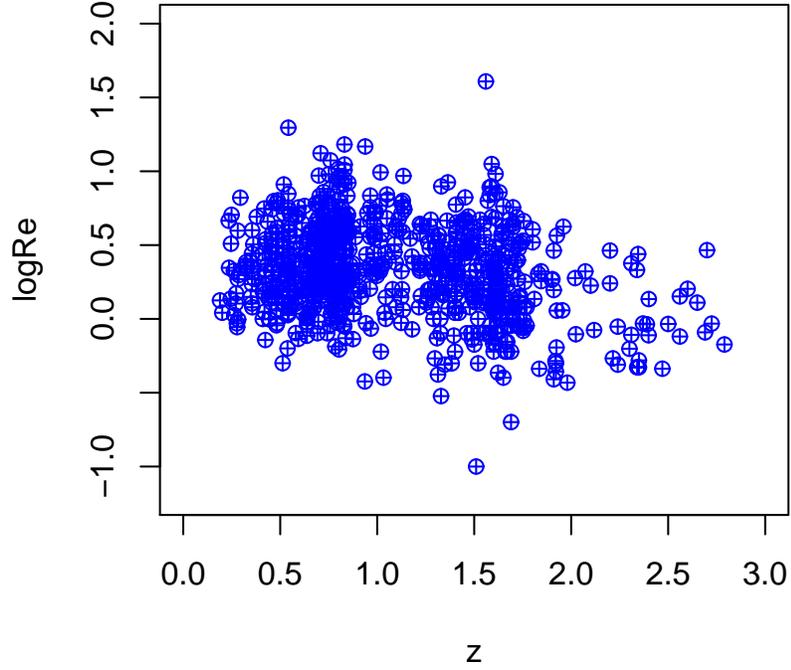}\\
  \caption{Logarithm of the effective radius versus redshift plot for the entire sample of ETGs in $0.2 \leq z \leq 2.7$}
  \label{Figure1}
\end{figure}

\begin{figure}
  \centering
  \includegraphics[width=0.5\textwidth]{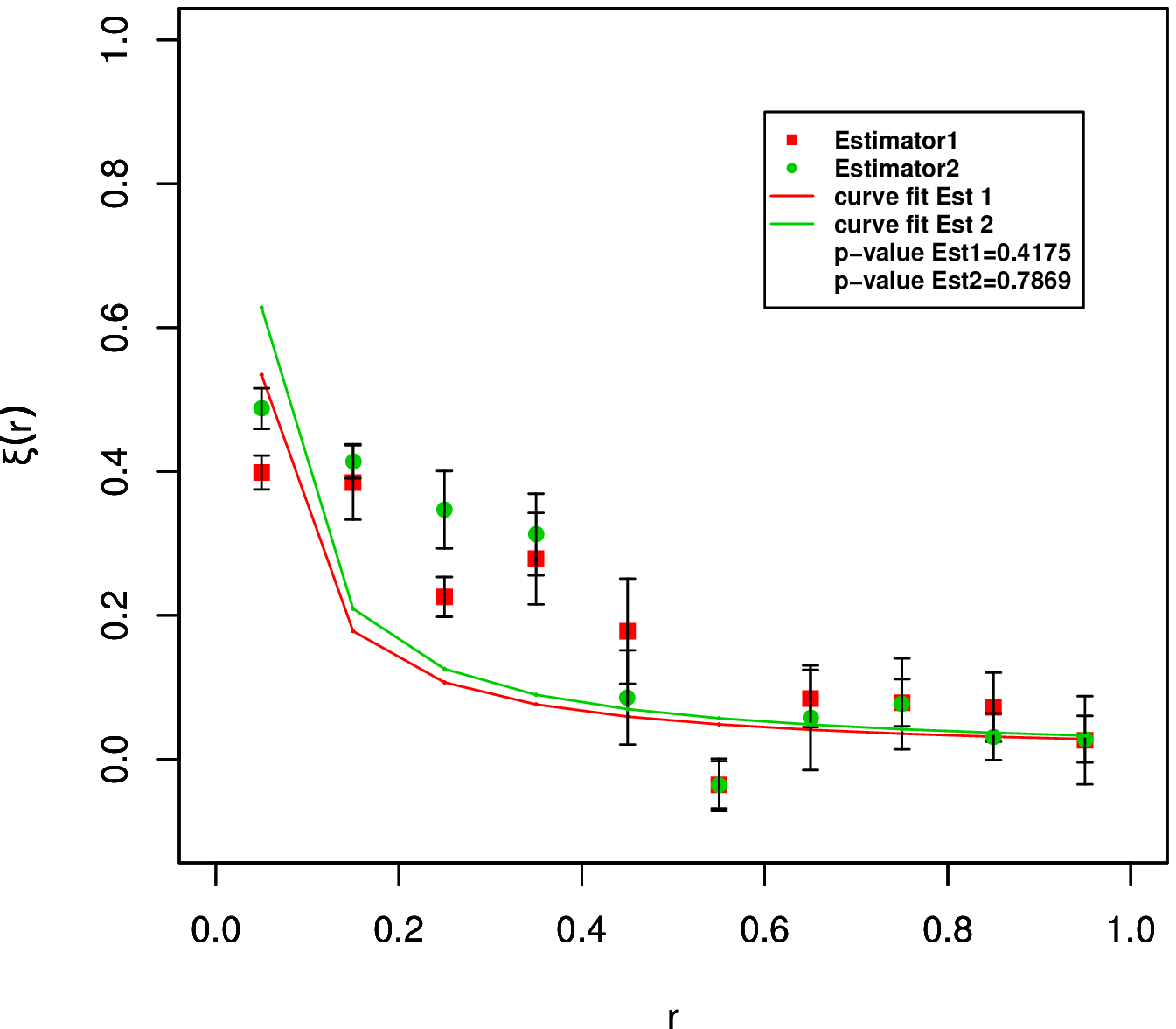}\\
  \caption{Cross-correlation function $\xi(r)$ versus normalized distance bin r between data sets 1 and 8. The solid lines are power
  laws for both the estimates as $\xi(r) \propto \frac{1}{r}$}
  \label{Figure2}
\end{figure}

\newpage
\begin{figure}
  \centering
  \includegraphics[width=0.5\textwidth]{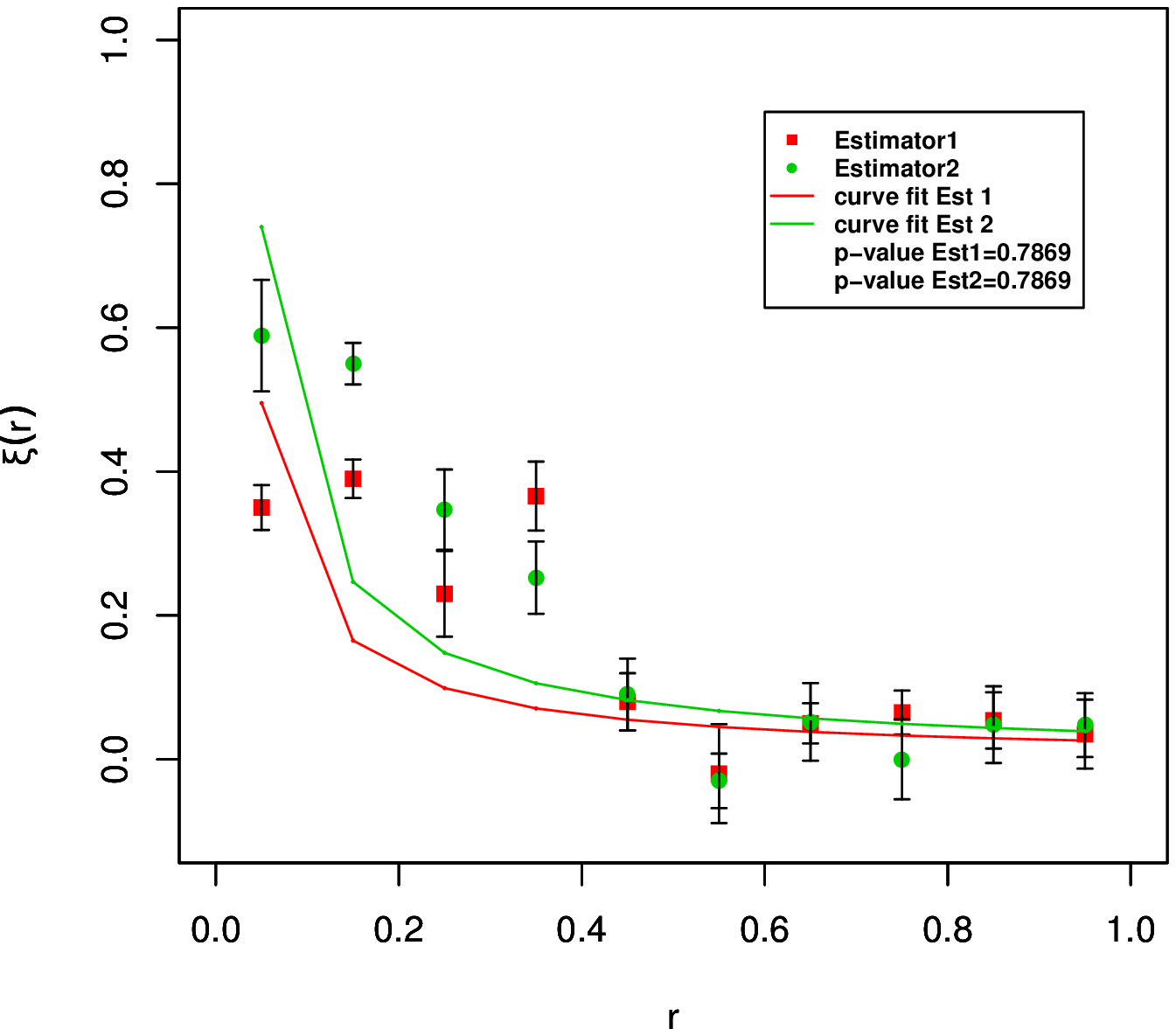}\\
  \caption{Cross-correlation function $\xi(r)$ versus normalized distance bin r between data sets 2 and 4.
  The solid lines are power law fits for both the estimates as $\xi(r) \propto
  \frac{1}{r}$}
  \label{Figure3}
\end{figure}

\newpage
\begin{figure}
  \centering
  \includegraphics[width=0.5\textwidth]{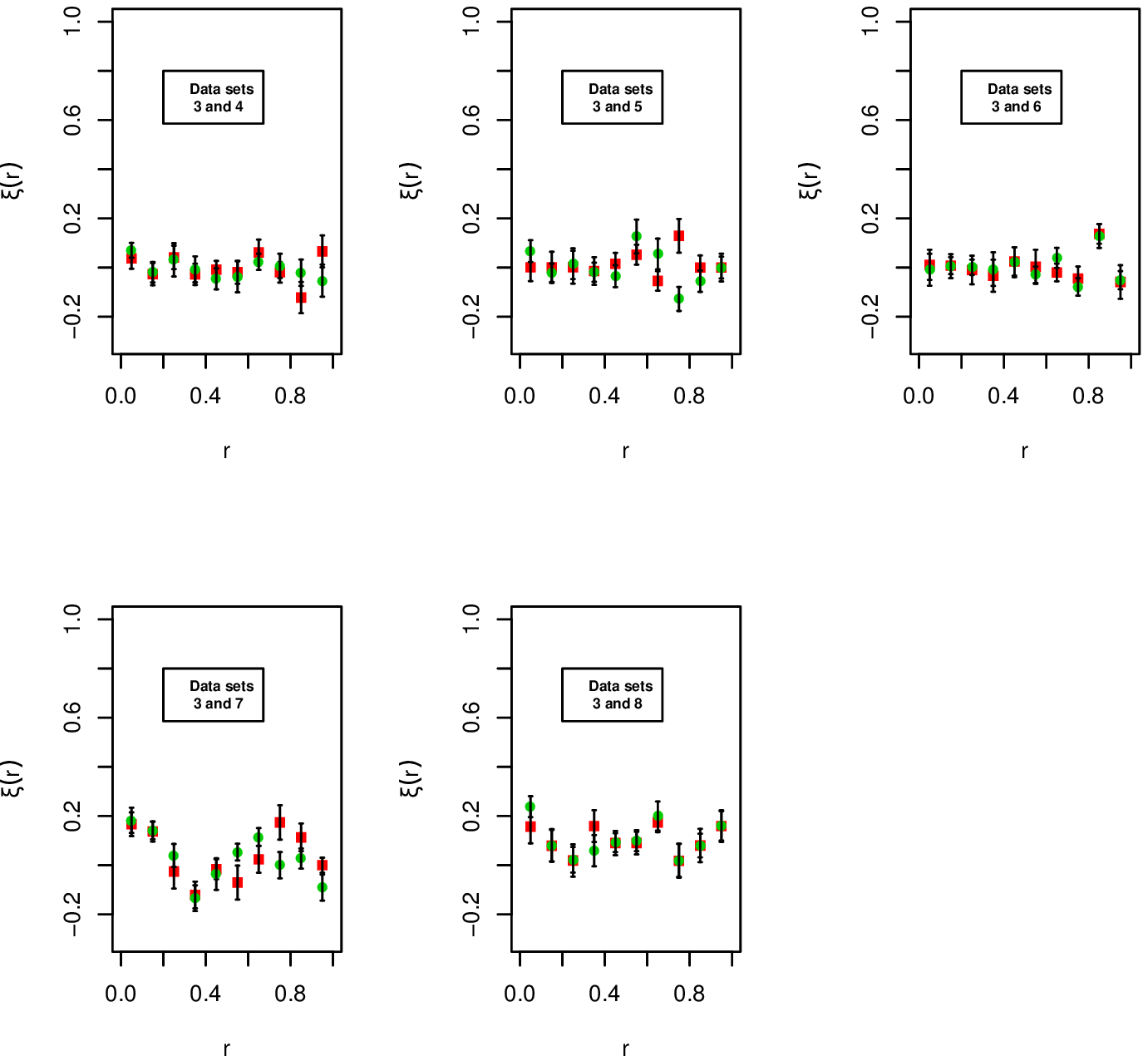}\\
  \caption{Cross-correlation function $\xi(r)$ versus normalized distance bin r between data set 3 and all redshift bins}
  \label{Figure4}
\end{figure}

\section{Appendix}
Multivariate multisample matching test:\\

For studying the compatibility among data under a multivariate set
up, the equality of location measures (mean, median, mode etc.) and
dispersion measures (sd, range etc) are of interest. If the joint
distribution of the parameters under consideration is multivariate
normal then MANOVA test is appropriate otherwise use of a
multivariate two sample nonparametric method is a better option. A
short description of the multivariate
non parametric test used is given below.\\

Let
$${X_{\alpha}^{(k)} = (X_{1\alpha}^{(k)},...,X_{p\alpha}^{(k)})',
\alpha = 1,...,n_{k},k = 1,...,c}$$ be
 a set of independent vector-valued random values, where c is the total
 number of populations, $n_{k}$ is the sample size of the kth population
 and p is the total number of parameters. The cumulative distribution
 function (c.d.f.) of $X_{\alpha}^{(k)}$ is denoted by $F_{k}(x)$.
 The set of admissible hypotheses designates that each $F_{k}(x)$ belongs to same class
 of distribution functions $\Omega$. The hypothesis to be tested, say $H_{0}$, specifies
 that $$H_0: F_{1}(x)=...=F_{c}(x)=F(x),  \forall  x, where F \in \Omega.$$
 The alternative to $H_{0}$ is the hypothesis that each $F_{k}(x)$ belongs to
 $\Omega$
 but that $H_{0}$ does not hold. To avoid the problem of ties, it is assumed that the
 class $\Omega$ is the class of all continuous distribution functions.\\ Here we pay particular
 attention to translation-type alternatives. For translation-type
 alternatives, we let $$F_{k}(x)=F(x+\delta_{k}),  \forall \ k = 1,...,c,  F \epsilon \Omega,$$ and
 we are interested in testing  (the reversed null hypothesis) $$H_{0}^{1} : \delta_{1} = ... = \delta_{c} = 0$$
 against the alternative that $\delta_{1}, ..., \delta_{c}$ are not all
 equal. \\
 We use the $``$Basic Rank Permutation Principle" given by Puri \& Sen (1970).
 Let us rank the N i-variate observations
 $X_{i\alpha}^{(k)}$, $\alpha$ = 1,...,$n_{k}$, k = 1,...,c in ascending order of magnitude,
 and let $R_{i\alpha}^{(k)}$ denote the rank of $X_{i\alpha}^{(k)}$ in this set.
 The observation vector $X_{\alpha}^{(k)}=(X_{1\alpha}^{(k)},...,X_{p\alpha}^{(k)})'$ then
 gives rise to the rank vector $R_{\alpha}^{(k)} = (R_{1\alpha}^{(k)},...,R_{p\alpha}^{(k)})'$,
 $\alpha$=1,...,$n_{k}$, k=1,...,c. The N rank vectors corresponding to the N observation
 vectors, $N=n_1+n_2+....+n_c$, can be represented by the rank matrix
 $$R_{N}^{p\times N} =
    \left(
      \begin{array}{ccccccccc}
        R_{11}^{(1)} & . & . & . & R_{1n_{1}}^{(1)} & . & . & . & R_{1n_{c}}^{(c)} \\
        . & . & . & . & . & . & . & . & . \\
        . & . & . & . & . & . & . & . & . \\
        . & . & . & . & . & . & . & . & . \\
        R_{p1}^{(1)} & . & . & . & R_{pn_{1}}^{(1)} & . & . & . & R_{pn_{c}}^{(c)} \\
      \end{array}
    \right)$$
Each row of this matrix is a random permutation of the numbers
1,2,...,N. Thus, $R_{N}^{p\times N}$ is a random matrix which can
have $(N!)^{p}$ possible realizations. Two rank matrices of the
above form are said to be permutationally equivalent if one can be
obtained from the other by a rearrangement of its columns. Thus a
matrix $R_N$ is permutationally equivalent to another matrix
$R_N^*$ which has the same column vectors as in $R_N$ but they are
so arranged that the first row of $R_N^*$ consists of the numbers
1,2,....N in the natural order i.e. $$R_N^{* p\times N}=
 \left(
      \begin{array}{cccccc}
        1&2&.&.&.&N\\
        R_{21}^{*} & . & . & . &.&R_{2N}^* \\
        . & . & . & . & . & . \\
        . & . & . & . & . & .  \\
        . & . & . & . & . & . \\
        R_{p1}^{*} & . & . & . & .&R_{pN}^* \\
      \end{array}
    \right)$$
  \\

In order to perform a $``$Permutation Rank Order Test" we start with
a general class of rank scores defined by explicitly known functions
of the ranks 1,...,N, viz.,
$$E_{N,\alpha}^{(i)} = (\frac{\alpha}{N+1}),$$$$1\leq\alpha\leq N,i=1,...,p.$$
Now, replacing the ranks $R_{i\alpha}^{(k)}$ in $R_{N}$ by
$E_{N,R_{i\alpha}^{(k)}}^{(i)}$, for all i = 1,...,p, $\alpha$ =
1,...,$n_{k}$, k=1,...,c, we get a corresponding p $\times$ N
matrix of general scores, which we denote by $E_{N}$. Thus,
$$E_{N} = \left(
                          \begin{array}{ccccccccc}
                            E_{N,R_{11}^{(1)}}^{(1)} & . & . & E_{N,R_{1n_{1}}^{(1)}}^{(1)} & . & . & E_{N,R_{1n_{c}}^{(c)}}^{(1)} \\
                            . & . & . & . & . & . & . \\
                            . & . & . & . & . & . & . \\
                            E_{N,R_{p1}^{(1)}}^{(p)} & . & . & E_{N,R_{pn_{1}}^{(1)}}^{(p)} & . & . & E_{N,R_{pn_{c}}^{(c)}}^{(p)} \\
                          \end{array}
                        \right)
$$We then consider the average rank scores for each i (=1,...,p) of the c samples, defined by
$$T_{N_{i}}^{(k)}=\frac{1}{n_{k}}\sum_{\alpha=1}^{n_{k}}E_{N,R_{i\alpha}^{(k)}}^{(i)},k=1,...,c, i=1,...,p.$$
Then, by straightforward computations,
$$v_{ij}(R_{N}^{*})=\frac{1}{N}\sum_{k=1}^{c}\sum_{\alpha=1}^{n_{k}}E_{N\alpha,i}^{(k)}E_{N\alpha,j}^{(q)}-\overline{E}_{N}^{(i)}\overline{E}_{N}^{(j)}$$
can be obtained, where $E_{N\alpha,i}^{(k)}$ is the value of
$E_{N,S}^{(i)}$ associated with the rank S = $R_{i\alpha}^{(k)}$,
and
$$\overline{E}_{N}^{(i)}=\sum_{\alpha=1}^{N}E_{N,\alpha}^{(i)}/N,i=1,...,p,$$
Now denoting $$V(R_{N}^{*})=((v_{ij}(R_{N}^{*})))_{i,j=1,...,p},$$
and following the structure of the test asymptotically equivalent
to chi-square statistic, we take as our test statistic
$\pounds_{N}$,$$
\pounds_{N}=\sum_{k=1}^{c}n_{k}[(T_{N}^{(k)}-\overline{E}_{N})V^{-1}(R_{N}^{*})(T_{N}^{(k)}-\overline{E}_{N})']$$
where $V^{-1}(R_{N}^{*})$=$((v_{ij}(R_{N}^{*})))^{-1}$,
$T_{N}^{(k)}$=($T_{N1}^{(k)}$,...,$T_{Np}^{(k)}$) and
$\overline{E}_{N}$=($\overline{E}_{N}^{(1)}$,...,$\overline{E}_{N}^{(p)}$).
\\ Let p = number of parameters, c = number of populations,\\
N = $\sum_{i=1}^{c}$$n_{i}$, $n_{i}$ is the sample size of the ith
sample,\\
i=1,...,c, $m_{H}$ = c-1, $m_{E}$ = n-c. The statistic
$\pounds_{N}$
can be approximated by $m_{E}$cF,\\
where F follows F distribution with a,b degrees of freedom.\\
Here a = p$m_{H}$, b = 4+$\frac{a+2}{B-1}$, c =
$\frac{a(b-2)}{b(m_{E}-p-1)}$, where
B = $\frac{(m_{E}+m_{H}-p-1)(m_{E}-1)}{(m_{E}-p-3)(m_{E}-p)}$.\\
This approximation was done by McKeon (1974). In order to compute
the value of the statistic we have used R - code.


\clearpage
\newpage

\begin{thebibliography}{00}
\bibitem{Arimoto}
Arimoto,N. and Yoshi,Y.(1987)," Chemical and photometric properties of a galactic wind model for elliptical galaxies," Astronomy and Astrophysics, 173,23-38
\bibitem{Ashman}
Ashman,K.M. $\&$ Zept,S.E.(1992), "The Formation of Globular Clusters in Merging and. Interacting Galaxies", Astrophysical Journal,384,50-61
\bibitem{Bell}
Bell,E.F.and de Zong R.S.(2001),"Stellar Mass-to-Light Ratios and the Tully-Fisher Relation", Astrophysical Journal, 550, 212-229
\bibitem{Bernardi}
Bernardi, M., Roche, N., Shankar, F.and Sheth, R. K.(2011),"Evidence of major dry mergers at $M_{*}>2 X 10^{11}M_{\bigodot}$ from curvature in early-type galaxy scaling relations?", Monthly Notices of Royal Astronomical Society,412, L6-L10
\bibitem{Bezanson}
Bezanson,R., van Dokkum,P.G., Tal,T., Marchesini,D., Kriek, M.,
Franx, M. and Coppi, P.(2009),"The Relation Between Compact, Quiescent High-redshift Galaxies and Massive Nearby Elliptical Galaxies: Evidence for Hierarchical, Inside-Out Growth",Astrophysical Journal, 697(2), 1290-1298
\bibitem{Blake}
Blake,C., Pope,A., Scott,D.and Mobasher,B.(2006),"On the cross-correlation of sub-mm sources and optically selected galaxies", Monthly Notices of Royal Astronomical Society, 368, 732-740
\bibitem{Bluck}
Bluck, A. F. L., Conselice, C. J., Buitrago, F., et al.(2012),"The Structures and Total (Minor + Major) Merger Histories of Massive Galaxies up to $z \sim 3$ in the HST GOODS NICMOS Survey: A Possible Solution to the Size Evolution Problem", Astrophysical Journal,747, 34
\bibitem{Bok}
Bok,B.J.(1934),Bull Harvard Observation, 895,1
\bibitem{Cappellari}
Cappellari, M., di Serego Alighieri, S., Cimatti, A., et al.(2009),"Dynamical Masses of Early-Type Galaxies at $z \sim 2$: Are they Truly Superdense?",Astrophysical Journal, 704, L34
\bibitem{Carlberg}
Carlberg,R.G.(1984),"Dissipative formation of an elliptical galaxy", Astrophysical Journal, 286, 403-415
\bibitem{Chandrasekhar}
Chandrasekhar,S. and Munch,G. (1952),"The theory of fluctuations in brightness of the Milky Way", Astrophysical Journal,115,103-123
\bibitem{Cimatti}
Cimatti,A., Cassata,P., Pozzetti,L.et al.(2008),"GMASS ultradeep spectroscopy of galaxies at $z \sim 2$. II. Superdense passive galaxies: how did they form and evolve?",Astronomy and Astrophysics, 482(1) ,21-42
\bibitem{Chattopadhyay}
Chattopadhyay, A.K., Mondal, S. and Chattopadhyay,T.(2013),"Independent Component Analysis for the objective classification of globular clusters of the galaxy NGC 5128", Computational Statistics \& Data Analysis",57,17-32
\bibitem{Chattopadhyay}
Chattopadhyay, A.K.,Chattopadhyay, T., Davoust, E., Mondal,
S.Sharina, M.(2009),"Study of NGC 5128 Globular Clusters Under Multivariate Statistical Paradigm", Astrophysical Journal, 705, 1533-1547
\bibitem{Conselice}
Conselice, C.J., et al.(2011),"The Hubble Space Telescope GOODS NICMOS Survey: overview and the evolution of massive galaxies at $1.5<z<3$",Mon.Not.Roy.Astron.Soc.,413,80-100.
\bibitem{Daddi}
Daddi, E., Renzini, A., Pirzkal, N., et al.(2005),"Passively Evolving Early-Type Galaxies at $1.4<z<2.5$ in the Hubble Ultra Deep Field", Astrophysical Journal, 626, 680-697
\bibitem{Damjanov}
Damjanov, I., Abraham, R. G., Glazebrook, K., et al.(2011),"Red nuggets at high redshift: structural evolution of quiescent galaxies over 10 Gyr of Cosmic history",  Astrophysical Journal, 739, L44-L50
\bibitem{Dekel}
Dekel, A., Sari, R., Ceverino, D.(2009),"Formation of massive galaxies at high redshift: cold streams,clumpy disks, and compact spheroids", Astrophysical Journal, 703, 785-801
\bibitem{DeLucia}
De Lucia,G.,Blaizot,J.(2007),"The hierarchical formation of the brightest cluster galaxies", Monthly Notices of Royal Astronomical Society, 375(1),2-14
\bibitem{Feldmann}
Feldmann,R., Carollo,C.M., Mayer,L.(2011),"The Hubble sequence in groups: the birth of the early-type galaxies" Astrophysical Journal, 736(2), 11-22
\bibitem{Ferre}
Ferr$\acute{e}$-Mateu, A., Vazdekis,A., Trujillo, I.et al.(2012),"Young ages and other intriguing properties of massive compact galaxies in the local Universe", Monthly Notices of Royal Astronomical Society, 423(1),632-646
\bibitem{Forbes}
Forbes,D.A., Bordie,J.P., \& Grillmair,C.J.(1997),"On the origin of Globular Clusters in Elliptical and cD Galaxies",Astronomical Journal,113,1652-1660
\bibitem{Grogin}
Grogin, N. A., Kocevski, D. D., Faber, S. M., et al.(2011),"Candles: the cosmic assembly near-infrared deep extragalactic legacy survey", Astrophysical Journal Suppliment,,197, 35
\bibitem{Guo}
Guo, Q.,White, S. D. M.(2008),"Galaxy growth in the concordance $\Lambda $CDM cosmology", Monthly Notices of Royal Astronomical Society,384(1),2-10
\bibitem{Ho}
Ho,L.C.,Li,Z.u.,Barth,A.J.,Seigar,M. S., Peng,C.Y., 2011, Astrophysical Journal, 197, 21-54
\bibitem{Hopkins}
Hopkins,P.F, Carton, D., Bundy,K. et al. (2010),"Mergers in $\Lambda$CDM : Uncertainties in Theoretical Predictions and interpretations of the Merger Rate", Astrophysical Journal, 724, 915-945
\bibitem{Huang}
Huang, S., Ho, L. C., Peng, C. Y.et al. (2013),"The Carnegie-Irvine Galaxy Survey. III. The Three-component Structure of Nearby Elliptical Galaxies",Astrophysical Journal,,766,47
\bibitem{Huang}
Huang, S., Ho, L. C., Peng, C. Y.et al. (2013a),"Fossil Evidence for the Two-phase Formation of Elliptical Galaxies",
Astrophysical Journal Letters, 768,L28-L33
\bibitem{Johansson}
Johansson, P. H., Naab, T.,  Ostriker, J. P.,(2012),"Forming Early-type Galaxies in $\Lambda $CDM Simulations. I. Assembly Histories", Astrophysical Journal,, 754, 115-137
\bibitem{Khochfar}
Khochfar,S.and Silk,J.(2006),"A simple model for the size evolution of Elliptical Galaxies", Astrophysical Journal Letters,648(1),L21-L24
\bibitem{Khochfar}
Khochfar,S.and Silk,J.(2006a)," On the origin of Stars in bulges and Elliptical Galaxies",Monthly Notices of Royal Astronomical Society ,370(2),902-910
\bibitem{Kormendy}
Kormendy,J.,Fisher,D.B.,Cornell,M.E.,Bender,R.(2009),"Structure and Formation of Elliptical and Spheroidal Galaxies", Astrophysical Journal Suppliments,,182(1),216-309
\bibitem{Larson}
Larson,R.B.(1975),"Models for the formation of elliptical galaxies", Monthly Notices of Royal Astronomical Society, 173,671-699
\bibitem{Limber}
Limber, D.N.(1953), "The analysis of counts of the extragalactic nebulae in terms of a fluctuating density field", Astrophysical Journal, 117,134
\bibitem{Limber}
Limber, D.N.(1954),"The analysis of counts of the extragalactic nebulae in terms of a fluctuating density field II",Astrophysical Journal, 119,655
\bibitem{Martinez}
Martinez, V.J. and Saar E.(2002),"Statistics of the galaxy distribution",Chapman \& Hall.
\bibitem{Mckean}
McKean,D.C.,Biedermann,S.,B$\ddot{u}$ger,H.(1974),"CH bond lengths and strengths, unperturbed CH stretching frequencies, from partial deuteration infrared studies: t-Butyl compounds and propane",Spectrochimica
Acta Part A: Molecular Spectroscopy 30(3),845-857
\bibitem{McLure}
McLure,R.J.,Pearce,H.J.,Dunlop,J.S. et al.(2012), Monthly Notices of Royal Astronomical Society, submitted(arXive:1205.4058)
\bibitem{Mihos}
Mihos, J.C.,Harding, P.,Spengler, C. E., Rudick, C. S., Feldmeier, J.
J.(2013),"The Extended Optical Disk of M101", Astrophysical Journal, 762(2), 82-95
\bibitem{Mondal}
Mondal,S., Chattopadhyay,T.,Chattopadhyay,A.K.,(2008),"Globular Clusters in the Milky Way and Dwarf Galaxies: A Distribution-Free Statistical Comparison", Astrophysical Journal,683,172-177
\bibitem{Mowbray}
Mowbray, A.G.(1938),"Non-random distribution of Extragalactic nabulae",PASP,50,275
\bibitem{Naab2009}
Naab,T.,Johansson, P.H.,Ostriker,J.P.(2009),"Minor Mergers and the Size Evolution of Elliptical Galaxies", Astrophysical Journal Letters,699(2),L178-L182
\bibitem{Naab2013}
Naab,T.(2013),"Modelling the formation of today's massive ellipticals",IAU Symposium no.295,Thomas D. Pasquali and A.Q.Ferreras Eds
\bibitem{Newman}
Newman, A. B., Ellis, R. S., Bundy, K., Treu, T.(2012),"Can Minor Merging Account for the Size Growth of Quiescent Galaxies? New Results from the CANDELS Survey ", Astrophysical Journal, 746,162
\bibitem{Neyman1956}
Neyman,J.,Scott,E.L.,Shane,C.D., 1956, Proceedings of the Third
Berkeley Symposium on Mathematical Statistics and Probability,
252,75-111
\bibitem{Neyman1954}
Neyman,J.,Scott,E.L.,Shane,C.D.(1954),"The Index of Clumpiness of the Distribution of Images of Galaxies",Astrophysical Journal Suppliment, 1, 269
\bibitem{Neyman1952}
Neyman,J.,Scott,E.L.(1952),"A Theory of the Spatial Distribution of Galaxies",Astrophysical Journal, 116,144
\bibitem{Nilsson}
Nilsson, M. K. M.,Klausen, M. B.,Söderlund, U., Ernst, R. E.(2013),"Precise U-Pb ages and geochemistry of Palaeoproterozoic mafic dykes from southern West Greenland: Linking the North Atlantic and the Dharwar cratons",
LITHOS, 174, 255-270
\bibitem{Onodera}
Onodera, M., Renzini, A., Carollo, M., et al.(2012),"Deep Near-infrared Spectroscopy of Passively Evolving Galaxies at z $>$1.4", Astrophysical Journal, 755, 26
\bibitem{Oser2010}
Oser, L., Ostriker, J. P., Naab, T., Johansson, P. H.,  Burkert, A.(2010),"The Two Phases of Galaxy Formation",Astrophysical Journal,725, 2312-2323
\bibitem{Oser2012}
Oser, L., Naab, T.,Ostriker, J.P., Johansson, P.H.(2012),"The Cosmological Size and Velocity Dispersion Evolution of Massive Early-type Galaxies" Astrophysical Journal,744(1), 63-72
\bibitem{Papovich}
Papovich, C., Bassett, R., Lotz, J. M., et al.(2012),"CANDELS Observations of the Structural Properties of Cluster Galaxies at z = 1.62", Astrophysical Journal, 750, 93-107
\bibitem{Peebles1974}
Peebles, P. J. E., Hauser, M. G.(1974),"Statistical Analysis of Catalogs of Extragalactic Objects. III. The Shane-Wirtanen and Zwicky Catalogs ", Astrophysical Journal Suppliment,28,19
\bibitem{Peebles}
Peebles, P. J. E.(1980),"The large scale structure of the Universe", Princeton University Press,U.K.
\bibitem{Prieto}
Prieto, M., Eliche-Moral, M. C., Balcells, M., et al.(2013),"Evolutionary paths among different red galaxy types at 0.3 < z < 1.5 and the late buildup of massive E-S0s through major mergers", Monthly Notices of Royal Astronomical Society, 428, 999-1019
\bibitem{Puri}
Puri,M.L. and Sen, P.K.(1966),"On a class of Multivariate Multisample rank-order tests",Sankhya A, 28(4),353-376
\bibitem{Salpeter}
Salpeter, E. E. (1955),"The Luminosity Function and Stellar Evolution.", Astrophysical Journal, 121, 161
\bibitem{Saracco}
Saracco,P.,Longhetti,M., Gargiulo,A.(2011),"Constraining the star formation and the assembly histories of normal and compact early-type galaxies at $1<z<2$", Monthly Notices of Royal Astronomical Society, 412,2707-2716
\bibitem{Sersic}
Sersic,J.L.(1968),"Atlas de Galaxies Australes",Cordoba:Observatoria Astronomico
\bibitem{Szomoru}
Szomoru, D., Franx, M.,  van Dokkum, P. G.(2012),"Sizes and Surface Brightness Profiles of Quiescent Galaxies at $z \sim 2$", Astrophysical Journal, 749, 121-132
\bibitem{Toomre}
Toomre, A.,  Toomre, J.(1972),"Galactic Bridges and Tails", Astrophysical Journal, 178, 623-666
\bibitem{Trujilo}
Trujillo, I., Feulner, G., Goranova, Y., et al.(2006),"Extremely compact massive galaxies at z ~ 1.4", Monthly Notices of Royal Astronomical Society, 373,L36-L40
\bibitem{Van Dokkum}
van Dokkum,P.G.(2008),"Evidence of Cosmic Evolution of the Stellar Initial Mass Function", Astrophysical Journal, 674(1),29-50
\bibitem{Van Dokkum}
van Dokkum, P. G., Whitaker, K. E., Brammer, G., et al. (2010),"The Growth of Massive Galaxies Since z = 2", Astrophysical Journal,709, 1018-1041
\bibitem{Whitaker}
Whitaker,K.E.,van Dokkum,P.G.,Brammer,G.,Franx,M.,(2012),"The Star Formation Mass Sequence Out to z = 2.5", Astrophysical Journal Letters,754(2),L29-L34
\bibitem{Zept}
Zept,S.E. et al.(2000),"Dynamical Constraints on the formation of NGC 4472 and its Globular Clusters", Astronomical Journal, 120,2928-2937
\bibitem{Zwicky}
Zwicky, F.(1953),"Supernovae", Helv Phys Acta, 26,241-254
\end{thebibliography}
\end{document}